\documentclass[%
reprint, showpacs,
amsmath,amssymb,
aps,
prapp,
]{revtex4-2}
\usepackage[utf8]{inputenc}
\usepackage{latexsym}
\usepackage{amssymb}
\usepackage{amsmath}
\usepackage{amsfonts}
\usepackage{mathtools}
\usepackage{cancel}
\usepackage{graphicx}
\usepackage{titlesec}
\usepackage{xcolor}
\usepackage{lipsum}
\usepackage{natbib}
\usepackage[T1]{fontenc}
\usepackage[utf8]{inputenc}
\usepackage{url}
\usepackage{appendix}
\titleformat{\paragraph}
{\normalfont\normalsize\bfseries}{\theparagraph}{1em}{}
\titlespacing*{\paragraph}
{0pt}{3.25ex plus 1ex minus .2ex}{1.5ex plus .2ex}

\begin{document}

\title{\large \textbf{Layer-Peeling Inversion of Temporal Refractive Index Profiles from Noisy Data}}


\author{Ohad Silbiger}
\affiliation{School of Electrical and Computer Engineering, Tel Aviv University, Tel Aviv 69978, Israel}

\author{Yakir Hadad}
\affiliation{School of Electrical and Computer Engineering, Tel Aviv University, Tel Aviv 69978, Israel}
\affiliation{Department of Electrical and Computer Engineering, The Ohio State University, Columbus, Ohio 43210, USA}
\email{yakirhad@tauex.tau.ac.il}

\date{\today}

\begin{abstract}

We introduce a direct method for reconstructing the temporal evolution of a medium from electromagnetic field measurements. A sequence of temporal refractive-index changes splits an incident wave into forward- and backward-propagating components, producing a \emph{spatial} response of finite extent. We show that this finite support, together with causality, enables sequential reconstruction of the temporal interfaces through a time-domain extension of layer peeling. For continuously varying media, we introduce a time-adjusted Goupillaud medium and, through space-time duality, establish the corresponding momentum-conservation and admissibility conditions. The method remains robust under incomplete and noisy measurements, providing a general framework for the metrology, characterization, and design of time-varying electromagnetic systems.

\end{abstract}

\maketitle

\section{Introduction} 
The interaction of electromagnetic (EM) waves with time-varying media has been studied for decades \cite{Morgenthaler1958, Cassedy1963, Felsen1970, Fante1971, Yu2026}. Such media have been shown to support a variety of unconventional wave phenomena, including nonreciprocity, temporal bandgaps, harmonic generation, and photonic time-crystal effects \cite{Hadad2015, Hadad2016, Lustig2018, Shaltout2019, Ozlu2025, Sustaeta2026}. In recent years, technological advances have made it increasingly feasible to move from theoretical concepts toward practical implementations and to harness temporal modulation for the manipulation of EM waves \cite{Lira2012, Estep2014, Taravati2017, Jaffray2026, Li2026}. This progress has been enabled, in large part, by the ability to rapidly modulate electromagnetic devices and metasurfaces on time scales sufficiently short to substantially affect high-frequency electromagnetic fields \cite{Reiskarimian2016, Lustig2023, Tirole2023}.

In parallel, inverse scattering has long played an important role in electromagnetics, providing a framework for inferring the properties of electromagnetic systems from field measurements \cite{Bruckstein1987, Colton1996, VanDenBerg1997}. Closely related inverse-design approaches address the complementary problem of determining the system properties required to produce a prescribed field response \cite{Burger2004, Moleski2018}.

The convergence of these two research directions has led to growing interest in inverse problems for time-varying media, and particularly in determining the temporal variation of material properties required to realize a desired wave response \cite{Ramaccia2021, Castaldi2021, SilbigerHadad2023, Galiffi2023}. A powerful conceptual tool for addressing such problems is space-time duality \cite{Agrawal2025}, through which models and techniques developed for spatially varying media can be adapted to systems that vary in time. This correspondence has enabled the temporal counterparts of numerous established concepts and methods, including scattering matrices \cite{Ramaccia2021, Xu2022}, Fresnel coefficients \cite{Xiao2014}, layered-media formulations \cite{ShlivinskiHadad2018, ValdezGarcia2024, Pacheco2025}, finite-difference time-domain methods \cite{Taravati2024}, transmission-line and input-impedance descriptions \cite{SilbigerHadad2023}, and quarter-wave transformers \cite{Castaldi2021}.

Inverse design, however, assumes that the desired response is prescribed and seeks a temporal modulation that realizes it \cite{Ramaccia2021, SilbigerHadad2025, Castaldi2026}. The corresponding \emph{metrology} problem is fundamentally different: the temporal evolution of the medium is unknown and must be inferred from measured electromagnetic fields. In practice, these measurements are inevitably incomplete and contaminated by noise, making direct application of inverse-design methodologies difficult. In this work, we show that causality, together with the finite support of the \emph{spatial impulse response}, provides sufficient structure to enable robust reconstruction of temporally varying material properties from such measurements. Specifically, we develop a time-domain extension of layer peeling that sequentially reconstructs the temporal interfaces and extend the formulation to continuously varying media through a time-adjusted Goupillaud model. The resulting framework enables accurate reconstruction even from incomplete and noisy data, providing a direct approach to the metrology and characterization of time-varying electromagnetic systems.

%
%
\section{Description of the system}
We consider an infinite and homogeneous, non-dispersive medium, with permittivity $\epsilon_0$ and permeability $\mu_0$. Throughout the analysis, the permeability remains constant, so the material can be characterized by a single parameter - the refractive index $n$. Inside the medium, a TEM wave $E_0^+,H_0^+$ with a finite width in space is propagating to the right (in the positive $x$ direction). At time $t=0$, the refractive index of the medium is switched instantaneously, to $n_1$. Due to the continuity of magnetic flux and electric displacement before and after switching, reflections will be caused. The reflection and transmission coefficients are given by \cite{ShlivinskiHadad2018}:
\begin{equation} \label{eq:rt_coeffs}
    R_1 = \frac{1}{2}\frac{n_0}{n_1}\left( \frac{n_0}{n_1} - 1 \right) \,\,\,\,
    T_1 = \frac{1}{2}\frac{n_0}{n_1}\left( \frac{n_0}{n_1} + 1 \right) .
\end{equation}
For simplicity of discussion, from now on only the electric field will be addressed. After switching the medium properties from $n_0$ to $n_1$, a right traveling wave and a left traveling wave will be propagating in the medium: $E_0(x, t) = TE_0^+(x - c_1 t) + R E_0^+(x + c_1 t)$, where $c_1 = c_0/n_1$ is the speed of light after switching ($c_0$ is the speed of light in vacuum). More generally, we may assume a superposition of right and left traveling waves is propagating in the medium when switching of the properties occurs, and the waves are allowed to propagate for a period of $\Delta t_1$ after the change. The relationship between the waves before and after this procedure (a temporal layer) is given by the following matrix equation \cite{Ramaccia2021}:
\begin{equation} \label{eq:switching_matrix}
\small
\setlength{\arraycolsep}{3pt}
\begin{bmatrix}
E_1^+ \\ E_1^-
\end{bmatrix} =
\begin{bmatrix} -\Delta x_1 &  0\\ 0 & \Delta x_1 \end{bmatrix}
\begin{bmatrix} T_1 & R_1 \\ R_1 & T_1 \end{bmatrix}
\begin{bmatrix} E_0^+ \\ E_0^- \end{bmatrix} \triangleq  
Q_1 \begin{bmatrix} E_0^+ \\ E_0^- \end{bmatrix} ,
\end{equation}
where the $\Delta x_1$ operator is a shift operator in space: $\Delta x_1 E = E(x + \Delta x_1) = E(x + c_1 \Delta t_1)$.

Now let's consider that this process is repeated $N$ times, with unknown values $\{n_i\}_{i=1}^N$ and propagation times $\{\Delta t_i\}_{i=1}^N$. In the following, we outline a method to solve the inverse scattering problem using this scheme, i.e. reconstruct the refractivity values and time durations, by knowledge of the input signal at time $t=0$ and the output signal (at time $t=\sum_{i=1}^N \Delta t_i$), as well as the state of the medium at the end - $n_N$. Without loss of generality, we will assume that at $t=0$ only a right-traveling wave exists in the medium. Observing Eq.~(\ref{eq:switching_matrix}), the overall field after the series of switching events is: 
$\mathbf{E}_N=Q_NQ_{N-1}...Q_1\mathbf{E}_0$, where $\mathbf{E}_i=[E^+_i, E^-_i]^T$, and $Q_i$ is the transfer operator of switching $i$, representing the switching from $n_{i-1}$ to $n_i$ and propagation over time $\Delta t_i$. This idea is conveyed in the \emph{spatial} impulse response of this system, which will be finite for any finite $N$ (FIR). Since the system is linear and space invariant, the output response can be represented as a convolution of the input field with a spatial impulse response $h(x)$: $E^+_{N} = h * E_0^+$. Here $h$ is the impulse response for transmission, but the same is true for reflection. Fig. \ref{fig:concept}(a)+(b) shows the evolution of the temporal (IIR) and spatial (FIR) impulse responses for structures with spatial and temporal layers, respectively. {The total number of propagation paths is $2^N$, distributed between the forward- and backward-propagating responses; generically, each contains $2^{N-1}$ path contributions.} We will show in this paper, that the impulse response finiteness, together with causality, are the fundamental concepts that allow the reconstruction of the medium temporal variation from field measurements in $N$ steps.
%
%
%

\section{The deconvolution process}
The layer peeling method is a well known method to directly and efficiently recover the properties of a layered material from the reflection response \cite{Bruckstein1987}. The key idea underlying this method is causality, which implies the response of deeper layers is delayed relative to the response of the more superficial ones. Thus, for example, if the reflection is measured as a function of time at the entrance to a transmission line, there exists a certain time $t_1$, that for all $t<t_1$ the measured reflection is caused only by the first interface. This fact allows reconstruction of the line's properties. Then, we may 'peel' this layer (once its properties are known) and apply the same {principle} on the second layer, and so on. Even though there is an infinite number of reflections, the peeling allows to recover the material properties in $N$ steps, $N$ being the number of interfaces in the line. Here we apply similar, yet somewhat different considerations, to recover the properties of the material in time. 

\begin{figure*}
    \includegraphics[width=\textwidth]{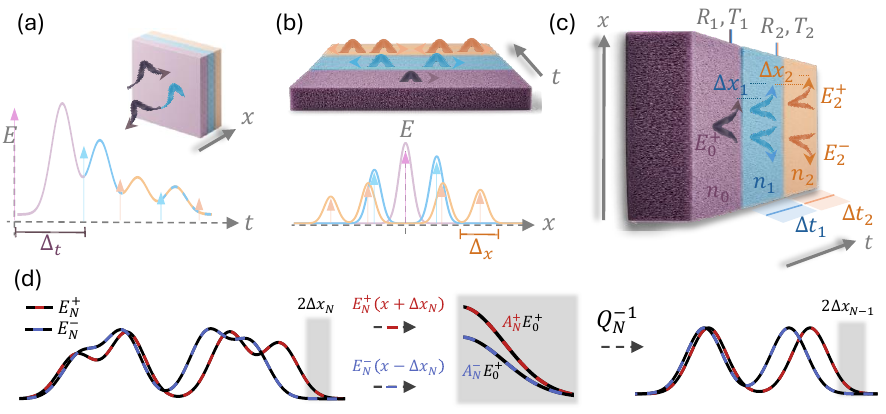}
    \centering
    \caption{Evolution of the impulse response in layered systems. (a) In space, causality indicates that reflections from superficial layers will appear {before reflections} from deeper ones. This idea renders that there is a certain duration $\Delta_t$ \emph{at the beginning of the reflected signal} through which there is no signal overlap, and the response is related to the incident wave only through the properties of the first interface. This is the main layer peeling concept that allows recovering the layer parameters. (b) In time, each property switching separates the signal into two waves (right and left traveling), where waves that are further away from $x=0$ underwent less direction inversions. Causality indicates there is a certain length $\Delta_x$ \emph{at the end of the reflected/transmitted signal} through which there is no signal overlap. In both cases, this part of the signal is related to the incident wave through multiplication by a single constant, which allows the recovery of the first layer (in space) or the last layer (in time). (c) A space-time illustration of the temporal layer model in discussion, with relevant notations for the analysis. For layer $i$, the refractive index of the medium is $n_i$, for duration $\Delta t_i$. The distance each wave has propagated in space during this time is $\Delta x_i$ (d) An illustration of the temporal layer peeling algorithm. \textit{Left} - the non-overlapping region $2\Delta x_N$ can be found from the supports of $E_{N}^+$ and $E_{N}^-$. \textit{Center} - the signals are shifted in space {and} the properties of the last switching are found. \textit{Right} - the signals are shifted back in time through inversion $Q_N^{-1}$, thus reducing their supports, and the process can be repeated.}
    \label{fig:concept}
\end{figure*}

Returning to the medium undergoing temporal switching, based on the discussion above it is clear that the right and left traveling waves after $N$ switching events - $E^+_N(x)$ and $E^-_N(x)$, will be a superposition of the {input field} $E_0^+(x)$ (i.e. the field at time $t=0$), shifted to various locations in space, equivalent to convolution with $h(x)$. To portray a clear picture of the properties of $E^+_{N}(x)$ and $E^-_{N}(x)$ that will reveal the medium parameters in earlier times, let us consider $E^+_{N-1}(x)$. In this state (as in any other), there exists a wave component in $E^+_{N-1}(x)$ that was never reflected, and has always propagated to the right. Of all the wave components, this component has traveled the longest distance to the right, which can be expressed as: $x_N = \sum_{i=1}^{N} \Delta x_i = \sum_{i=1}^{N} c_{i}\Delta t_i$. $\Delta x_i$ denotes the distance each of the waves traveled between switching events $i$ and $i+1$. See an illustration of these parameters in Fig~\ref{fig:concept}(c), for a system with $N=2$. Any other component (right traveling or left traveling), \emph{should be to the left} of this wave. We note here that we are only considering the right side of the output response, however the same is true for the left side of the output response. Now the parameters are switched one more time, from $n_{N-1}$ to $n_{N}$, and the waves are allowed to propagate for time $\Delta t_N$. Now the pulse in discussion was separated into two components, as per Eq.~(\ref{eq:switching_matrix}). These arguments, as well as the fact that the input pulse is of finite length, lead to two important takeaways:
(1) There exists a point in space - $x_f$ such that $E^+_{N}(x) = 0 \,\, \forall \, x > x_f$.
(2) There exists a point in space to the left of $x_f$ (that we will denote by $x_p$), such that to its right, only one replica of the input wave exists (there is a region with no signal overlap). The first takeaway stems directly from the fact that the system is FIR, while the second takeaway stems from causality. Therefore, in the vicinity of the right edge of the signal (denoted by a gray square in Fig~\ref{fig:concept}(d)), the signal can be expressed as:
%
%
\begin{subequations}\label{eq:vright_final}
\small
\begin{align}
\!\! E^+_{N}(x) &=
\begin{cases}
A^+_N E_0^+(x-x_N), & x_p < x < x_f,\\
0, & x > x_f,
\end{cases}\\
\!\! E^-_{N}(x) &=
\begin{cases}
A^-_N E_0^+(x - x_N + 2\Delta x_N),
& \begin{aligned}[t]
x_p &< x+2\Delta x_N < x_f ,
\end{aligned}\\
0, & x+2\Delta x_N > x_f .
\end{cases}
\end{align}
\end{subequations}
In the equations above, $A^+_N = \Pi _{i=1}^N T_i$ is the amplitude of the rightmost right-traveling wave and $A^-_N = \Pi _{i=1}^{N-1} T_i R_N$ is the amplitude of the rightmost left-traveling wave. By knowledge of the supports of $E^+_{N}$ and $E^-_{N}$ we may find $\Delta x_N$, and shift the signals accordingly, to the state in which they were $\Delta t_N$ time units ago, right after switching. Then, using Eq.~(\ref{eq:vright_final}) together with Eq.~(\ref{eq:rt_coeffs}), the refractive index before switching - $n_{N-1}$, is recovered.
%
%
%
%
The last step is to shift the signals back in time, to the state in which they were before the switching, and then repeat the process. This can be done easily by inverting Eq.~(\ref{eq:switching_matrix}):
%
%

\begin{equation} \label{eq:inverse_matrix}
\small
\setlength{\arraycolsep}{2pt}
\!\!\! \begin{bmatrix}
E_{N-1}^+(x) \\ E_{N-1}^-(x)
\end{bmatrix}
=
\frac{1}{T_N^2 - R_N^2}
\begin{bmatrix}
T_N & -R_N \\
-R_N & T_N
\end{bmatrix}
\begin{bmatrix}
E_N^+(x+\Delta x_N) \\
E_N^-(x-\Delta x_N)
\end{bmatrix}.
\end{equation}
The last step reduces the number of nonzero terms of $h(x)$ by a factor of 2, since the new overlap between the wave components will cancel out half of them. This allows the process to be over by $N$ steps. Because this is in-fact a shift of the waves back in time, the takeaways that were mentioned above are valid for $E_{N-1}^+$ and $E_{N-1}^-$, as they were for $E_N^+$ and $E_N^-$, and the process can be repeated. To summarize the algorithm: 
    (1) Find the supports of $E_N^+$ and $E_N^-$ (on the right side of the $x$ axis), denoted by $x_f^+$ and $x_f^-$.
    (2) Find the spatial shift between the waves: $\Delta x_N = \frac{1}{2} (x_f^+ - x_f^-)$. This promptly gives $\Delta t_N$.
    (3) Shift the signals $E_N^+$ and $E_N^-$ and recover $n_{N-1}$ using Eqs.~(\ref{eq:rt_coeffs}) and (\ref{eq:vright_final}) in the vicinity of $x_f^+$ and $x_f^-$.
    (4) Shift the signals back in time using Eq.~(\ref{eq:switching_matrix}) and return to (1).
    The process is terminated when the reflected wave {vanishes}.

{
To demonstrate the applicability of the general layer-peeling procedure to
abrupt temporal variations, we next consider a refractive-index profile
consisting of several successive discontinuities whose temporal locations are
not known a priori. In this case, we use the general formulation described
above, rather than the equal-traveling-distance discretization introduced in
Sec.~IV. At each peeling step, the propagation distance associated with the
last temporal interval is inferred from the relative locations of the edges of
the forward- and backward-propagating signals, after which the corresponding
temporal interface is recovered and removed. Figure~\ref{fig:discontinuity}
shows an example in which a sequence of sharp changes in $n(t)$ is accurately
reconstructed from the measured fields. A finite-support trapezoidal incident
waveform is used so that the relevant signal edges can be identified
accurately. This example also illustrates an important limitation of the
general procedure: errors in determining the signal edges accumulate through
successive peeling steps. Consequently, reconstruction of abrupt
discontinuities at unknown temporal locations becomes increasingly sensitive
to noise and edge-localization errors as the number of temporal interfaces
increases.

\begin{figure}[t]
    \centering
    \includegraphics[width=\columnwidth]{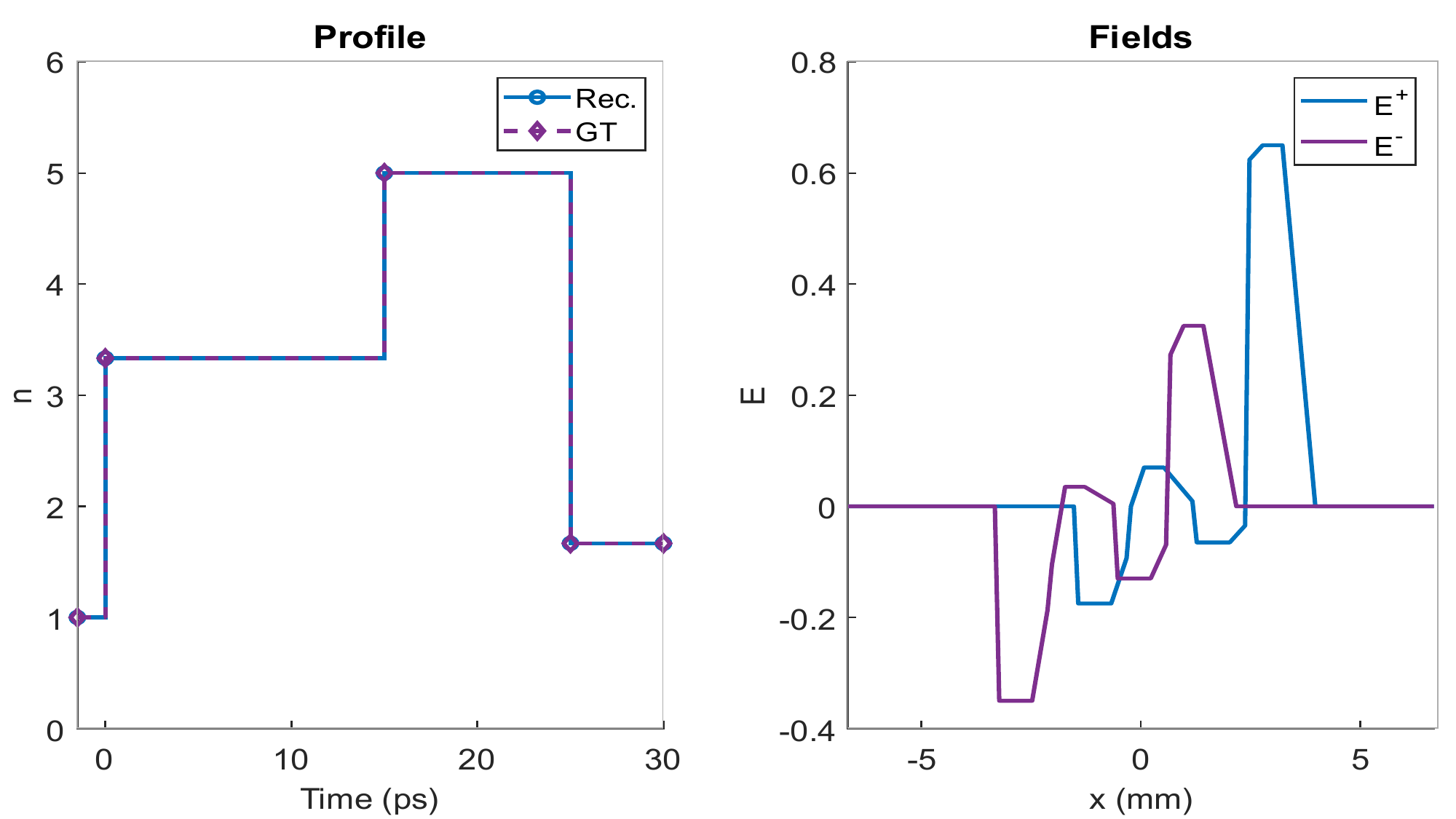}
    \caption{{
    Reconstruction of a temporal refractive-index profile containing abrupt
    discontinuities using the general layer-peeling procedure of Sec.~III.
    (a) Ground-truth (labeled GT in the legend) refractive-index profile $n(t)$ and its reconstruction.
    The temporal locations of the discontinuities are not assumed to be known
    a priori, but are recovered sequentially from the measured fields.
    {(b) Corresponding forward- and backward-propagating fields obtained using a finite-support trapezoidal incident waveform.
    The sharply defined signal edges facilitate accurate estimation of the propagation
    distance at each peeling step.} Such an accuracy is not required for the method suggested in Sec. V. There, a smooth Gaussian signal is used for demonstration. 
    }}
    \label{fig:discontinuity}
\end{figure}
}

%
%

\section{Equal traveling distance layers}
{The algorithm detailed in the previous section is exact for a piecewise-constant temporal profile and can approximate continuously varying profiles through sufficiently fine temporal discretization, for any input signal with finite support.} For the sake of concreteness, and since our ultimate goal is the practical {metrology} of continuously varying time-dependent media, we shall assume from now on that the distance traveled by the waves \emph{between two consecutive switching events} is constant, i.e. $\Delta x_i = \Delta x \,\,\,\forall i$. This choice is inspired by analogous problems in space, in which the \emph{traveling time} in each layer is assumed constant ({Goupillaud} medium) \cite{Goupillaud1961}. The constant traveling distance allows mathematically compact expressions for the wave functions, which comes to our aid when recovering the medium parameters, and is very suitable for continuous switching. This is because the degree of freedom in choosing $\Delta t_i$ allows $\Delta x_i$ to be constant.
For a constant traveling distance between switching events, the transfer matrix between temporal layers $i$ and $i+1$ for a harmonic wave with wavenumber $k$ obtains the form:
%
%
\begin{equation} \label{eq:switching_matrix_etdl}
\begin{bmatrix}
E_{i}^+ \\ E_{i}^- \end{bmatrix} =
\begin{bmatrix} T_ie^{-jk\Delta x} & R_ie^{-jk\Delta x} \\ 
                R_ie^{jk\Delta x} & T_ie^{jk\Delta x} \end{bmatrix}
\begin{bmatrix} E_{i-1}^+ \\ E_{i-1}^- \end{bmatrix} ,
\end{equation}
Due to the spatial homogeneity, the wavenumber $k$ is conserved in time \cite{Castaldi2026}.
%
%
If we denote: $z = e^{2jk\Delta x}$ and $\tilde R_i = \frac{1}{2}(\frac{n_{i-1}}{n_{i}}-1)$, this equation becomes:
\begin{equation} \label{eq:switching_matrix_etdl_z}
\begin{bmatrix}
E_{i}^+ \\ E_{i}^- \end{bmatrix} = \frac{n_{i-1}}{n_{i}} z^{1/2}
\begin{bmatrix} (1+\tilde R_i)z^{-1} & \tilde R_i z^{-1} \\ 
                \tilde R_i & 1+\tilde R_i \end{bmatrix}
\begin{bmatrix} E_{i-1}^+ \\ E_{i-1}^- \end{bmatrix} ,
\end{equation}
Here we harness the $z$-transform notation from digital signal processing to indicate a constant spatial shift, or a phase shift that is linear with $k$. The field waves after $N$ switching events are obtained by multiplication of the transfer matrices in Eq.~(\ref{eq:switching_matrix_etdl_z}), resulting in polynomials in $z$. The overall transfer matrix of $N$ events would be:
%
%
\begin{equation} \label{eq:switching_matrix_etdl_full}
\begin{bmatrix}
E_{N}^+ \\ E_{N}^-
\end{bmatrix}
= \frac{n_0}{n_{N}} z^{N/2}
\begin{bmatrix}
A(z) & B(z) \\ 
C(z) & D(z)
\end{bmatrix}
\begin{bmatrix}
E_0^+ \\ E_0^-
\end{bmatrix} ,
\end{equation}
where,
%
%
\begin{equation} \label{eq:dtot}
\begin{bmatrix}
A(z) & B(z) \\ 
C(z) & D(z)
\end{bmatrix} =
\prod_{i=1}^N \begin{bmatrix} (1+\tilde R_i)z^{-1} & \tilde R_i z^{-1} \\ 
                \tilde R_i & 1+\tilde R_i \end{bmatrix} .
\end{equation}
{The product is ordered so that the matrix corresponding to $i=1$ is rightmost.} The full transmission and reflection coefficients would be:
\begin{subequations}
\begin{align} \label{eq:tr_etdl}
    T_{\mbox{tot}}(k)      &= \frac{n_0}{n_{N}} z^{N/2} \, A(e^{2jk\Delta x}) \\
    \Gamma_{\mbox{tot}}(k) &= \frac{n_0}{n_{N}} z^{N/2} \, C(e^{2jk\Delta x}) .
\end{align}
\end{subequations}
Here, $A(z)$ is of degree $N$ and $C(z)$ is of degree $N-1$. The form of Eq.~(\ref{eq:dtot}) is familiar from scattering theory in space \cite{Orfanidis2016}. The transfer matrices in space and time can be related, thus providing conditions of admissibility. {For the real, positive refractive indices considered here,} space-time duality renders that $A$ has to be a \emph{maximum phase} system, i.e., all the zeros of $A$ must lie outside of the unit circle. Also, the conserved quantity in space invariant systems is momentum (the temporal equivalent of energy conservation) \cite{Yin2023}. This renders that on the unit circle:
\begin{equation} \label{eq:momentum_conservation}
    \frac{n_N}{n_0} |{A}(e^{j\theta})|^2 - \frac{n_N}{n_0}|{C}(e^{j\theta})|^2 = 1 .
\end{equation}
See App.~A and App.~B for derivations of these conditions. As shown in Eqs.~(\ref{eq:dtot}) and (\ref{eq:tr_etdl}), the transmission and reflection responses are represented through the $z$-transform of their impulse responses. Therefore, the coefficients may be calculated through the spectrum of $T_{\mbox{tot}}$ and $\Gamma_{\mbox{tot}}$, or directly from the spatial representation of the signal using an optimization approach. While for design problems the former may be a more direct method, for {metrology} problems both methods are applicable. We have chosen the latter.

For the inversion process, we start by assuming $M$ switching events, where the signal travels a distance of $\Delta x$ between two consecutive events. The distance $\Delta x$ may be chosen arbitrarily, although should be sufficiently small. However, the overall distance $M\Delta x$ should correspond to the actual distance traveled by the signals, which may be estimated by the support of the measured right traveling signal $E_{\mbox{out}}^+$. Then, assuming we have the input signal $E_0^+$ and the output signals $E_{\mbox{out}}^+$ and $E_{\mbox{out}}^-$ as vectors, sampled at $L$ points, we use the input signal to construct the shift matrices $S^A, S^C \in \mathbb{R}^{L\times{M}}$:
%
%
\begin{subequations} \label{eq:sm_matrix}
\begin{equation} \label{eq:sm_matrix_A}
S^A_{ij} = 
        E_{0}^+\left( x_i+(M-2j)\Delta x \right) ,
\end{equation}
\begin{equation} \label{eq:sm_matrix_C}
S^C_{ij} = 
        E_{0}^+\left( x_i+(M-2j+2)\Delta x \right) .
\end{equation}
\end{subequations}
That is, each column of $S^A$ and $S^C$ is a shifted version of the input $E_{0}^+$. {Since Eq.~(\ref{eq:tr_etdl}) contains the common factor $n_0/n_M$, we define the normalized output vectors $\overline{\mathbf E}_{\mathrm{out}}^{\pm}=(n_M/n_0)\mathbf E_{\mathrm{out}}^{\pm}$.} Then we solve:
%
%

\begin{subequations}\label{eq:ls}
\begin{equation}\label{eq:ls_a}
\widehat{\mathbf a}
=
\operatorname*{arg\,min}_{\mathbf a}
\left\|
{\overline{\mathbf E}_{\mathrm{out}}^{+}}
-
S^{A}\mathbf a
\right\|_{2}^{2}
\end{equation}

\begin{equation}\label{eq:ls_c}
\widehat{\mathbf c}
=
\operatorname*{arg\,min}_{\mathbf c}
\left\|
{\overline{\mathbf E}_{\mathrm{out}}^{-}}
-
S^{C}\mathbf c
\right\|_{2}^{2}
\end{equation}
\end{subequations}

and find the coefficient vectors $\mathbf{a}$ and $\mathbf{c}$. These vectors represent the polynomials $A(z)$ and $C(z)$ as follows: $A(z) = \sum_{i=1}^M a_i z^{-i}$, $C(z) = \sum_{i=0}^{M-1} c_i z^{-i}$. From Eq.~(\ref{eq:dtot}) we note that: $a_M = \prod_{i=1}^M (1 + \tilde R_i)$ and $c_{M-1} = \tilde R_M\prod_{i=1}^{M-1} (1 + \tilde R_i)$, which renders: 
\begin{equation} \label{eq:rn}
    {\tilde R_M = \frac{c_{M-1}}{a_M-c_{M-1}}} .
\end{equation}
This result is consistent with the main idea of the general algorithm, that the properties of the last switching are 'hidden' in the part of the wave that has propagated the longest distance (or accumulated the largest phase).
After recovering $\tilde R_M$, the temporal layer is peeled off by inversion of Eq.~(\ref{eq:switching_matrix_etdl_z}), resulting in new polynomials for $A(z)$ and $C(z)$, of degrees $M-1$ and $M-2$, respectively. Performing this process $M$ times will recover the properties of all layers.
%
%

\section{Results}
\begin{figure}[h]
    \includegraphics[width=0.48\textwidth]{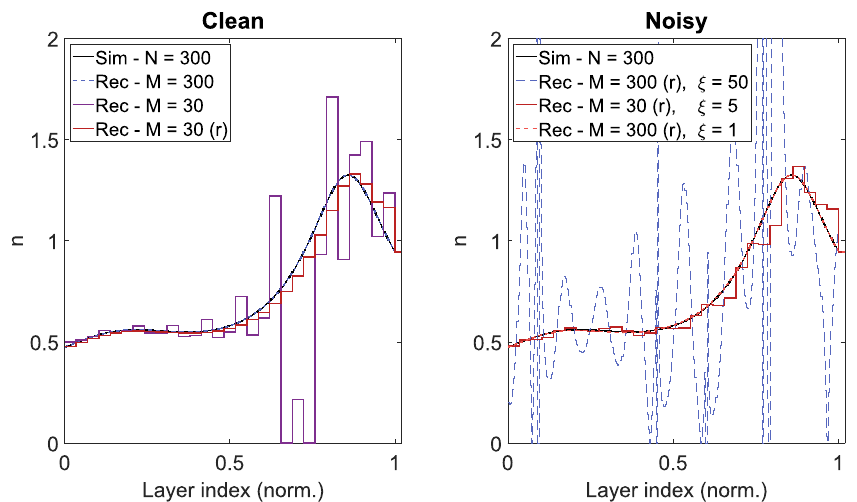}
    \centering
    \caption{Reconstruction of {refractive-index variation} in time from clean and noisy data (SNR = 15 dB). {(r) denotes reconstruction with regularization.}}
    \label{fig:results}
\end{figure}

This section demonstrates the suggested algorithm in action, for continuous modulation of the material properties. First, an arbitrary continuous modulation of $n(t)$ was simulated  with $N=300$. The resulting coefficients of $A(z)$ and $C(z)$ were calculated, and the measured signals $E_{\mbox{out}}^+(x)$ and $E_{\mbox{out}}^-(x)$ were generated numerically using these coefficients, assuming a Gaussian pulse input with spatial width (std) $\sigma_x$. Then, the output was used to reconstruct the properties of the layers, first by applying the least-squares (LS) method to solve {Eqs.~(\ref{eq:ls_a}) and~(\ref{eq:ls_c})}, and then recovering the values of $n$. An important key in the LS problem is the choice of the number of layers ($M$) for reconstruction. As shown in Fig.~\ref{fig:results}(left, blue curve), when no noise is present, and the algorithm assumes a number of layers which is comparable to the number of layers in the simulation ($M=N=300$), the results match the profile very well. Physically, this means that the layer thickness (in time) is small compared to the typical time of property variation, or, equivalently, that the assumption that $n(t)$ is constant throughout the layer duration $\Delta t_i$ is a good approximation of the continuous time variation. 
%
However, when the layer thickness is large compared to the actual variation time ($N=300, M=30$), the LS problem results in a large and discontinuous variation of the properties due to the need to fit a signal comprised of 300 shifted Gaussians using 30 shifted Gaussians. To overcome this issue, a regularization term was added to the LS problem:

\begin{subequations}\label{eq:ls_reg}
\begin{equation}\label{eq:ls_reg_a}
\widehat{\mathbf a}
=
\operatorname*{arg\,min}_{\mathbf a}
\left\{
\left\|
{\overline{\mathbf E}_{\mathrm{out}}^{+}}
-
S^{A}\mathbf a
\right\|_{2}^{2}
+
\lambda
\left\|
D\mathbf a
\right\|_{2}^{2}
\right\},
\end{equation}

\begin{equation}\label{eq:ls_reg_c}
\widehat{\mathbf c}
=
\operatorname*{arg\,min}_{\mathbf c}
\left\{
\left\|
{\overline{\mathbf E}_{\mathrm{out}}^{-}}
-
S^{C}\mathbf c
\right\|_{2}^{2}
+
\lambda
\left\|
D\mathbf c
\right\|_{2}^{2}
\right\},
\end{equation}
\end{subequations}
where $D$ is a second order difference matrix and $\lambda$ is the regularization parameter (we used $\lambda=0.001$). {The second-order difference penalty introduces a smoothness prior appropriate for the continuously varying profiles considered here and stabilizes the deconvolution when neighboring shifted replicas of the incident field become nearly linearly dependent. It therefore suppresses rapidly varying coefficient sequences that are weakly constrained by the measurements, reducing spurious oscillations in the reconstructed profile.} This significantly improves the reconstruction of the properties, as can be seen in Fig.~\ref{fig:results} (left, red curve).

With the inclusion of {additive white Gaussian noise (AWGN)} to the model, an additional issue must be addressed. We define the unitless parameter $\xi = \sigma_x/{\Delta x}$, which has a major effect on the recovery quality. In Fig.~\ref{fig:results} (right, blue curve), the pulse width $\sigma_x$ is large compared to the layer thickness ${\Delta x}$ ($\xi = 50$). In this case, reconstruction is not possible. This is because the columns of $S^A$ and $S^C$ are nearly dependent and the LS problem in Eqs.~(\ref{eq:ls}) becomes ill-conditioned. This can be solved by decreasing the number of layers (increasing ${\Delta x}$ - red curve) or by using a narrower input pulse (decreasing $\sigma_x$ - orange curve).

\begin{figure}[t]
    \centering    \includegraphics[width=\columnwidth]{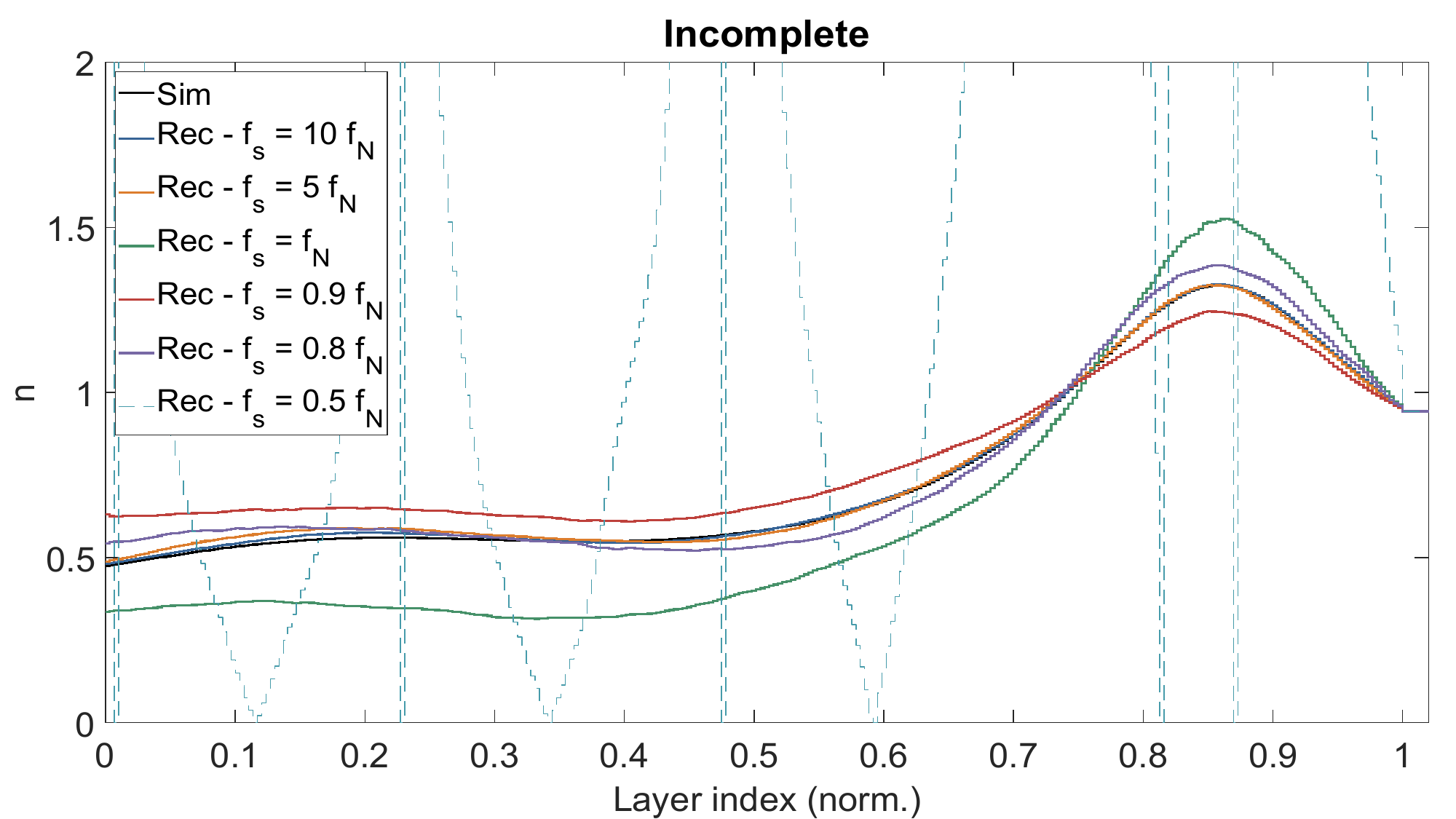}
    \caption{{Reconstruction of the temporal refractive-index
    profile from incompletely sampled output fields. The spatial sampling rate
    is varied relative to the reference Nyquist rate $f_{\mathrm{N}}$
    associated with the half-power bandwidth of the reflected signal.
    Measurement noise is included in all cases. Accurate reconstruction is
    maintained over a broad range of sampling rates, while at sufficiently low
    sampling rates the increased ill-conditioning and noise sensitivity lead
    to deterioration of the recovered profile. In the legend, `Sim' stands for the ground truth, while the next items labeled 'Rec' correspond to different levels of reconstruction incompleteness associated with the sampling rates {$10f_{\mathrm{N}},\,5f_{\mathrm{N}},\,f_{\mathrm{N}},\,0.9f_{\mathrm{N}},\,0.8f_{\mathrm{N}}$, and $0.5f_{\mathrm{N}}$}}}
    \label{fig:incomplete}
\end{figure}

{
To further examine reconstruction from incomplete measurements, we consider
finite-rate spatial sampling of the output fields. Since the reflected signal
is not strictly bandlimited, no finite sampling rate provides a complete
representation of its spatial spectrum. We therefore define an effective
bandwidth using the half-power bandwidth of the reflected signal, and denote
the corresponding Nyquist sampling rate by $f_{\mathrm{N}}$. Figure~\ref{fig:incomplete}
shows reconstruction results, including measurement noise, for sampling rates
ranging from $0.5f_{\mathrm{N}}$ to $10f_{\mathrm{N}}$. Reasonable
reconstruction is obtained over a broad range of sampling rates, including
rates close to and somewhat below $f_{\mathrm{N}}$, despite the incomplete
representation of the continuous spatial field. This robustness results from
the structured nature of the inverse problem together with the regularization
used in the reconstruction. As expected, however, reducing the sampling rate
increases the sensitivity to noise and to the sampling grid, and for the
lowest sampling rate considered, $0.5f_{\mathrm{N}}$, the reconstruction
becomes unstable.
}

\subsection{Scope and limitations.}
{
The present analysis considers finite, nonperiodic temporal modulations in a
lossless and nondispersive medium, with additive measurement noise applied to
the output fields. The transfer-matrix formulation remains formally applicable
to finite periodic modulations, including regimes exhibiting parametric gain;
however, strong amplification may substantially increase the dynamic range and
ill-conditioning of the forward operator, thereby increasing the sensitivity
of the reconstruction to noise. Noise already present in the incident field or
introduced during the modulation process may also undergo parametric
amplification. Material loss and dispersion introduce additional challenges:
loss reduces the available signal-to-noise ratio, whereas dispersion prevents
propagation between temporal interfaces from being represented by simple
spatial translations of the incident waveform. In principle, the framework may
be generalized to a prescribed causal dispersive model, such as a Drude or
Lorentz medium, by augmenting the electromagnetic state with the corresponding
material variables and formulating the peeling procedure in this enlarged state
space. Such extensions are expected to introduce additional conditioning and
identifiability challenges and are left for future work.
}


\section{Conclusions}
In this paper, we have established a robust and direct methodology for the reconstruction of temporal refractive index profiles from field measurements. By exploiting the unique FIR nature of the spatial impulse response, we have shown that the layer-peeling concept can be successfully adapted to provide direct, step-by-step inversion of temporal layers. The transition from a general piecewise model to Goupillaud medium allows the use of $z$-transform formalisms, enabling the recovery of continuous modulations through simple optimization methods. As technological capabilities in high-speed material modulation continue to evolve, the proposed temporal layer-peeling method offers a critical analytical tool for the characterization and diagnostic evaluation of devices with time varying properties. Moreover, the derived methods and conditions for admissibility can be readily used for the design of time-varying media. {
Extensions of the present framework to strongly amplifying, lossy, and
dispersive temporal media constitute natural directions for future work.
}



\appendix

\section{Spatial and Temporal Transfer Matrices}

In this section we present realizability conditions and the condition for conservation of momentum for temporal layers with equal traveling distance. This supplementary material will use a slightly different expression for the transfer matrix from the main text (up to a constant). This is to have a better vision of the (mathematical) duality between spatial-domain and time-domain transfer matrices.
\begin{figure} [t]
\includegraphics[width=\linewidth]{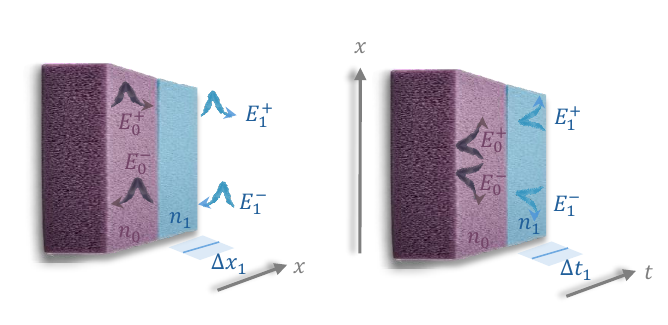}
    \centering
    \caption{Fields before and after a single spatial (left) and temporal (right) layer}
    \label{fig:layer}
\end{figure}

We consider a spatial layer as a transition between two materials, followed by propagation distance $\Delta x$ (see Fig.~\ref{fig:layer}-left). The transfer matrix of a single spatial layer reads:
\begin{align}
    \begin{bmatrix} \tilde{E}_0^+(t, x) \\ \tilde{E}_0^-(t, x) \end{bmatrix} 
    = \tilde{Q}_{01} 
    \begin{bmatrix} \tilde{E}_1^+(t, x + \Delta x) \\ \tilde{E}_1^-(t, x + \Delta x) \end{bmatrix} .
\end{align}
Here we adopt the notation of transfer from left to right, as in \cite{Orfanidis2016}. 
We consider a temporal layer as an instantaneous transition between two materials, followed by propagation time $\Delta t$ (see Fig.~\ref{fig:layer}-right). The transfer matrix of a single temporal layer reads:
\begin{align}
    \begin{bmatrix} \hat{E}_1^+(t + \Delta t, x) \\ \hat{E}_1^-(t + \Delta t, x) \end{bmatrix} 
    = \hat{Q}_{01} 
    \begin{bmatrix} \hat{E}_0^+(t, x)  \\ \hat{E}_0^-(t, x) \end{bmatrix} .
\end{align}
Henceforth, the tilde (\textasciitilde{}) denotes spatial-domain quantities, and the hat (\textasciicircum{}) will denote time-domain quantities. The waves are reflected by the interface (spatial/temporal), and are then allowed to propagate for some distance/time. The transfer matrix in space would be:
\begin{align}
\tilde{Q}_{01} &= \frac{1}{2n_0}
    \begin{bmatrix}
    (n_0 + n_1)  \, e^{jk_1\Delta x_1} & (n_0 - n_1) \, e^{-jk_1\Delta x_1} \\ 
    (n_0 - n_1)  \, e^{jk_1\Delta x_1} & (n_0 + n_1) \, e^{-jk_1\Delta x_1}
    \end{bmatrix} ,
\end{align}
and in time:
\begin{align} \label{eq:Qt_single}
\hat{Q}_{01} = \frac{n_0}{2n_1^2}
    \begin{bmatrix}
    (n_0 + n_1)  \, e^{-j\omega_1\Delta t_1} &(n_0 - n_1) \, e^{-j\omega_1\Delta t_1} \\ 
    (n_0 - n_1)  \, e^{j\omega_1\Delta t_1}  &(n_0 + n_1) \, e^{j\omega_1\Delta t_1}
    \end{bmatrix} .
\end{align}  
In the following, we will assume that there are multiple spatial layers, for which: $\Delta t_i = \Delta x_i/c_i = \Delta t$, i.e., the traveling time in each layer is constant. In an analogous fashion, we will assume that for multiple temporal layers, $\Delta x_i = \Delta t_i \cdot c_i = \Delta x$, i.e., the traveling \emph{distance} in each temporal layer is equal. Thus, we denote:
\begin{equation} \label{eq:z_definitions}
    \tilde{z} = e^{2j\omega \Delta t} \,\, , \,\, \hat{z} = e^{2jk \Delta x}
\end{equation}
Using this notation, the transfer matrix from any layer $i$ to layer $i+1$ becomes:
\begin{align} \label{eq:Qsi}
\tilde{Q}_{i,i+1} &= \frac{\tilde{z}^{1/2}}{2n_{i}}
    \begin{bmatrix}
    (n_{i} + n_{i+1}) & (n_{i} - n_{i+1})\tilde{z}^{-1} \\ 
    (n_{i} - n_{i+1}) & (n_{i} + n_{i+1})\tilde{z}^{-1}
    \end{bmatrix} \\
\hat{Q}_{i,i+1} &= \frac{\hat{z}^{1/2}n_{i}}{2n_{i+1}^2} \label{eq:Qti}
    \begin{bmatrix}
    &(n_{i} + n_{i+1})\hat{z}^{-1} &(n_{i} - n_{i+1})\hat{z}^{-1} \\ 
    &(n_{i} - n_{i+1})  &(n_{i} + n_{i+1})
    \end{bmatrix}
\end{align} 
The overall transfer matrix of $N$ interfaces ($N+1$ layers) in space and time would be:
\begin{equation} \label{eq:Q_products}
    \tilde{Q} = \prod_{i=0}^{N-1} \tilde{Q}_{i,i+1} , \,\,\, 
    \hat{Q} = \prod_{i=0}^{N-1} \hat{Q}_{i,i+1} ,
\end{equation}
so that:
\begin{subequations} \label{eq:QsQt_fields}
\begin{align}
    &\begin{bmatrix} 
        \tilde{E}_0^+(t, x) \\ 
        \tilde{E}_0^-(t, x) 
    \end{bmatrix} 
   =
    \tilde{Q}
    \begin{bmatrix} 
        \tilde{E}_N^+(t, x + \sum\Delta x_i) \\ 
        \tilde{E}_N^-(t, x + \sum\Delta x_i)
    \end{bmatrix},
    \label{eq:QsQt_fields_a}
    \\[4pt]
    &\begin{bmatrix} 
        \hat{E}_N^+(t + \sum\Delta t_i, x) \\ 
        \hat{E}_N^-(t + \sum\Delta t_i, x) 
    \end{bmatrix} 
    =
    \hat{Q}
    \begin{bmatrix} 
        \hat{E}_0^+(t, x) \\ 
        \hat{E}_0^-(t, x)
    \end{bmatrix}.
    \label{eq:QsQt_fields_b} 
\end{align}
\end{subequations}
The matrices can be written as:
\begin{align}
\tilde{Q} &= \frac{\tilde{z}^{N/2}\nu_S}{2^N} \prod_{i=0}^{N-1}
    \begin{bmatrix}
    (n_i + n_{i+1}) & (n_i - n_{i+1})\tilde{z}^{-1} \\ 
    (n_i - n_{i+1}) & (n_i + n_{i+1})\tilde{z}^{-1}
    \end{bmatrix} \label{eq:Qspace_expanded} \\
\hat{Q} &= \frac{\hat{z}^{N/2}\nu_T}{2^N} \prod_{i=0}^{N-1}
    \begin{bmatrix}
    &(n_i + n_{i+1})\hat{z}^{-1} &(n_i - n_{i+1})\hat{z}^{-1} \\ 
    &(n_i - n_{i+1})  &(n_i + n_{i+1})
    \end{bmatrix} \label{eq:Qtime_expanded}
\end{align} 
{where the multiplication is performed so that $\tilde{Q} =\tilde{Q}_{0,1}\cdot ...\cdot \tilde{Q}_{N-1,N}$ and $\hat{Q} =\hat{Q}_{N-1,N}\cdot ...\cdot \hat{Q}_{0,1}$, and}:
\begin{equation} \label{eq:nu_definitions}
    \nu_S = \prod_{i=0}^{N-1} \frac{1}{n_i} , \,\,\, 
    \nu_T = \prod_{i=0}^{N-1} \frac{n_i}{n_{i+1}^2}  .
\end{equation}

\section{Realizability condition for equal travel distance layers}

{For the dual discretizations, we choose $\Delta x=c\,\Delta t$, so that $\hat{z}(\omega/c)=\tilde{z}(\omega)$.}
From Eqs.~(\ref{eq:Qsi}) and (\ref{eq:Qti}) we see that:
\begin{equation} \label{eq:dual_single_layer}
    \hat{Q}_{i,i+1}(\omega/c) = \left(\frac{n_i}{n_{i+1}}\right)^2 J\tilde{Q}_{i,i+1}^T\left(\omega\right)J ,
\end{equation}
where we denoted 
$\hat{Q}\left(\omega/c\right) = \hat{Q}\left(e^{2j \frac{\omega}{c}\Delta x}\right)$ 
and 
$\tilde{Q}\left(\omega \right) = \tilde{Q}\left(e^{2j \omega\Delta t}\right)$. Also, {$c = c_0/n_0$}, $c_0$ being the speed of light in vacuum, and:
\begin{equation} \label{eq:J_matrix}
J = 
    \begin{bmatrix}
    0 & 1 \\ 
    1 & 0
    \end{bmatrix}
\end{equation}
Due to the fact that $JJ=I$, $I$ being the 2x2 identity matrix, we get:
\begin{equation} \label{eq:QsQt}
    \hat{Q}= \left(\frac{n_0}{n_{N}}\right)^2 J\tilde{Q}^TJ .
\end{equation}
Let's denote the entries of $\tilde{Q}$ and $\hat{Q}$ as per Eq.~(\ref{eq:switching_matrix_etdl_full}):
\begin{subequations} \label{eq:ABCD}
\begin{align}
    \tilde{Q}
    &=
    \left(\frac{n_0}{n_{N}}\right) \tilde{z}^{N/2}
    \begin{bmatrix}
        \tilde{A}(\tilde{z}) & \tilde{B}(\tilde{z}) \\ 
        \tilde{C}(\tilde{z}) & \tilde{D}(\tilde{z})
    \end{bmatrix},
    \label{eq:ABCD_a}
    \\[4pt]
    \hat{Q}
    &=
    \left(\frac{n_0}{n_{N}}\right) \hat{z}^{N/2}
    \begin{bmatrix}
        \hat{A}(\hat{z}) & \hat{B}(\hat{z}) \\ 
        \hat{C}(\hat{z}) & \hat{D}(\hat{z})
    \end{bmatrix}.
    \label{eq:ABCD_b}
\end{align}
\end{subequations}
Based on Eq.~(\ref{eq:QsQt}):
\begin{equation} \label{eq:AD_relation}
    \hat{D}\left(\omega/c\right) = 
    \left(\frac{n_0}{n_{N}}\right)^2 \tilde{A}\left(\omega\right)
\end{equation} 
From Eqs.~(\ref{eq:QsQt_fields}) and (\ref{eq:ABCD}), the reflection and transmission coefficients for the spatial and temporal $N$ layers are defined \cite{Orfanidis2016}:
\begin{equation} \label{eq:GammaT_space}
    \tilde{\Gamma}(\omega) = 
    \frac{\tilde{C}(\tilde{z} = e^{2j\omega \Delta t})}  
    {\tilde{A}(\tilde{z} = e^{2j\omega \Delta t})} , 
    \,\,\,\,\,\,\,\,\,
    \tilde{T}(\omega) = \frac{1}{\tilde{\gamma} \tilde{A}(\tilde{z} = e^{2j\omega \Delta t})} ,
\end{equation}
and:
\begin{equation} \label{eq:GammaT_time}
    \hat T(k) = \hat{\gamma}\hat{A}(\hat{z} = e^{2jk \Delta x}) , 
    \,\,\,\,\,\,\,\,\,
    \hat{\Gamma}(k) = \hat{\gamma}\hat{C}(\hat{z} = e^{2jk \Delta x}),
\end{equation}
{where for convenience we have denoted: $\tilde{\gamma} = \left(\frac{n_0}{n_{N}}\right) e^{j\omega N\Delta t}$ and $\hat{\gamma} = \left(\frac{n_0}{n_{N}}\right) e^{jk N\Delta x}$}. Realizability condition of the spatial layers is that the zeros of $\tilde{A}$ are inside the unit-circle, which guarantees the system is stable \cite{Orfanidis2016}. This also indicates that the spatial reflection coefficients (Fresnel coefficients) in each interface are smaller than 1 in absolute value (positive refraction values). An alternative definition is that $\tilde{A}$ is a minimum-phase system. Using Eqs.~(\ref{eq:AD_relation}), (\ref{eq:GammaT_space}) and (\ref{eq:GammaT_time}), we obtain the dual realizability condition for temporal layers, which is that $\hat{D}$ is a minimum-phase system. 
Equivalently, we may notice that $\hat{D}(\hat{z})$ is the reverse polynomial of $\hat{A}(\hat{z})$. This means that the vector representing the coefficients of $\hat{A}(\hat{z})$: $[a_N,...,a_0]$ is a flipped version of the vector representing the coefficients of $\hat{D}(\hat{z})$: $[d_N,...,d_0]$. This indicates that an equivalent condition would be that $\hat{A}(\hat{z})$ is maximum-phase.
Lastly, momentum conservation implies that $\forall k$ \cite{Yin2023}:
\begin{equation} \label{eq:momentum_conservation_appendix}
   \frac{n_N}{n_0} |\hat{A}(k)|^2 - \frac{n_N}{n_0}|\hat{C}(k)|^2 = 1
\end{equation}

\bibliography{bibliography} 

\end{document}